\documentclass[twocolumn,prb,showpacs,superscriptaddress,preprintnumbers,amsmath,amssymb,floatfix]{revtex4}
\usepackage{graphicx}
\usepackage{dcolumn}
\usepackage{bm}

\usepackage{ulem}
\usepackage{color}

\begin{document}


\title{Verwey transition in Fe$_{3}$O$_{4}$ thin films: Influence of oxygen stoichiometry and substrate-induced microstructure}

\author{X.~H.~Liu}
  \affiliation{Max Planck Institute for Chemical Physics of Solids, N\"othnitzerstr. 40, 01187 Dresden, Germany}
\author{A.~D.~Rata}
  \affiliation{Max Planck Institute for Chemical Physics of Solids, N\"othnitzerstr. 40, 01187 Dresden, Germany}
\author{C.~F.~Chang}
  \affiliation{Max Planck Institute for Chemical Physics of Solids, N\"othnitzerstr. 40, 01187 Dresden, Germany}
\author{A.~C.~Komarek}
  \affiliation{Max Planck Institute for Chemical Physics of Solids, N\"othnitzerstr. 40, 01187 Dresden, Germany}
\author{L.~H.~Tjeng}
  \affiliation{Max Planck Institute for Chemical Physics of Solids, N\"othnitzerstr. 40, 01187 Dresden, Germany}

\date{\today}

\begin{abstract}

We have carried out a systematic experimental investigation to address the question why thin films of Fe$_3$O$_4$ (magnetite) generally have a very broad  Verwey transition with lower transition temperatures as compared to the bulk. We observed using x-ray photoelectron spectroscopy, x-ray diffraction and resistivity measurements that the Verwey transition in thin films is drastically influenced not only by the oxygen stoichiometry but especially also by the substrate-induced microstructure. In particular, we found (1) that the transition temperature, the resistivity jump, and the conductivity gap of fully stoichiometric films greatly depends on the domain size, which increases gradually with increasing film thickness, (2) that the broadness of the transition scales with the width of the domain size distribution, and (3) that the hysteresis width is affected strongly by the presence of antiphase boundaries. Films grown on MgO (001) substrates showed the highest and sharpest transitions, with a 200 nm film having a \textit{T}$_V$ of 122~K, which is close to the bulk value. Films grown on substrates with large lattice constant mismatch revealed very broad transitions, and yet, all films show a transition with a hysteresis behavior, indicating that the transition is still first order rather than higher order.
\end{abstract}


\maketitle

\section{Introduction}

The theoretically predicted half-metallic character of Fe$_{3}$O$_{4}$ \cite{Groot83,Yanase84,Groot84,Groot86} has made magnetite one of the most studied transition metal oxide material for thin film applications in devices such as spin valves and magnetic tunnel junctions. A tremendous amount of work has been devoted to preparing thin films with high crystalline quality. Using a variety of deposition methods, epitaxial growth on a number of substrates has been achieved.\cite{Lind1992,Chambers1997,Chambers1997_2,Gao97,Chambers1997_4,Chambers1999,Kiyomura1999,Chambers2000,Chambers2000_2,ChambersRep2000,Voogt1997,Voogt1998,Voogt1999,
Bloemen1996,Gaines1997,Gaines1997_2,Anderson_Mg,Heijden95,Heijden1998,Fontijn1997,Hibma1999,Fuji1990,Fuji1990_2,Fuji1994,Fuji1999,Mijiritskii2001,Subagyo2006,Fonin2005,Moussy2013}

Yet, the physical properties of the thin films are not that well defined as those of the bulk material. In particular, the first order metal-insulator transition, known as Verwey transition \cite{Verwey39}, is in thin films very broad \cite{Eerenstein02,Kumar06,Sena97,Gong97,Kale01,Li98,Arora05,Arora06,Ramos06,Ziese00,Ziese2000,Ziese02,Geprags13,Wang13,Liu13,Naftalis11,Moussy04,Margulies96,Orna10} as compared to that in the bulk single crystal. The Verwey transition temperature $T_V$ in thin films is also much lower, with reported values ranging from 100 to 120~K\cite{Eerenstein02,Kumar06,Sena97,Gong97,Kale01,Li98,Arora05,Arora06,Ramos06,Ziese00,Ziese2000,Ziese02,Geprags13,Wang13,Liu13,Naftalis11,Moussy04,Margulies96,Orna10} while the stoichiometric bulk has $T_V$ of 124-125 K. It is not clear why the Verwey transition in thin films is so diffuse.

In this work, we investigate systematically the effect of oxygen stoichiometry, thickness, strain, and microstructure on the Verwey transition in epitaxial Fe$_3$O$_4$ thin films on a variety of substrates. We use molecular beam epitaxy (MBE) technique under ultra-high vacuum conditions combined with \textit{in situ} electron diffraction and spectroscopic characterization as well as \textit{ex situ} x-ray diffraction and electrical conductivity measurements. Our aim is to understand the factors that affect negatively the Verwey transition in thin films, so that we can identify the route to synthesize Fe$_3$O$_4$ thin films with transport properties as good as or perhaps even better than the bulk material.

\section{Experimental}

Fe$_{3}$O$_{4}$ thin films were grown by molecular beam epitaxy (MBE) in an ultra-high vacuum (UHV) chamber with a base pressure in the 1$\times$10$^{-10}$ mbar range. High purity Fe metal was evaporated from a LUXEL Radak effusion cell at temperatures of about $1250\,^{\circ}\mathrm{C}$ in a pure oxygen atmosphere onto single crystalline MgO (001), SrTiO$_3$ (001) (STO), and MgAl$_2$O$_4$ (001) (MAO) substrates. These substrates were annealed for 2 h at $600\,^{\circ}\mathrm{C}$ in an oxygen pressure of 3$\times$10$^{-7}$ mbar to obtain a clean and well-ordered surface structure prior to the Fe$_{3}$O$_{4}$ deposition. The substrate temperature was kept at $250\,^{\circ}\mathrm{C}$ during growth.

The Fe flux was calibrated using a quartz-crystal monitor at the growth position prior to deposition and set to 1~\textrm{\AA} per minute for the growth of all films. Molecular oxygen was simultaneously supplied through a leak valve. Fe$_{3}$O$_{4}$ films with different oxygen stoichiometries were grown by varying the partial oxygen pressure between 5$\times$10$^{-8}$ and 1$\times$10$^{-5}$ mbar, while keeping the Fe flux constant. The growth was terminated by simultaneously closing the oxygen leak valve and the Fe shutter.

\textit{In situ} and \textit{real-time} monitoring of the epitaxial growth was performed by reflection high-energy electron diffraction (RHEED) measurements. Oscillations in the RHEED specular beam intensity, where each oscillation corresponds to the formation of one new atomic monolayer (ML), allows for precise control of the film thickness. The crystalline structure was also verified \textit{in situ} after the growth by low-energy electron diffraction (LEED). All samples were also analyzed \textit{in situ} by x-ray photoelectron spectroscopy (XPS). The XPS data were collected using 1486.6 eV photons (monochromatized Al $K_{\alpha}$ light) in normal emission geometry and at room temperature using a Scienta R3000 electron energy analyzer. The overall energy resolution was set to about 0.3 eV.

Temperature dependent resistivity measurements of the Fe$_{3}$O$_{4}$ thin films were performed by standard four probe technique using a physical property measurement system (PPMS). X-ray diffraction (XRD) was employed for further \textit{ex situ} investigation of the structural quality and the microstructure of the films. The XRD measurements were performed with a high resolution Philips XPert MRD diffractometer using monochromatic Cu $K_{\alpha 1}$  radiation ($\lambda$ = 1.54056 \AA).

\section{Oxygen stoichiometry}

It is well known that the oxygen stoichiometry greatly influences the Verwey transition in bulk Fe$_{3(1-\delta)}$O$_4$ \cite{Aragon85,Kakol92,Aragon93,Honig95,Brabers98}. The order parameter shows a clear discontinuity across the transition in samples with -0.0005 $<$ $\delta$ $<$ ${\delta_c}$ = 0.0039, whereas for $\delta_c$ $\leq$ $\delta$ $\leq$ 3${\delta_c}$, the discontinuity disappears and the term second or higher order has been used to describe the temperature behavior of the order parameter. With increasing $\delta$, the transition becomes broad and the transition temperature, \textit{T}$_V$, decreases continuously.\cite{Aragon85,Aragon93,Brabers98} For thin films, on the other hand, the few reports available on the effect of oxygen stoichiometry are mostly focused on the crystal structure, morphology, magnetic properties, and resistivity at room temperature \cite{Chambers1997,Gao97,Kiyomura1999,VoogtPhD,Fuji1999,Huang2013}. In particular, little has been done to study the influence of the oxygen content on the Verwey transition itself with the resistivity as the order parameter in magnetite thin films.

To start our study, we first investigate the effect of the oxygen stoichiometry on the Verwey transition of single crystalline epitaxial Fe$_3$O$_4$ thin films. We prepared a series of samples, all with a thickness of 40~nm, where the Fe flux is set at 1~\textrm{\AA} /min and the substrate temperature at $250\,^{\circ}\mathrm{C}$. We varied the oxygen pressure in a wide range, from 5$\times$10$^{-8}$ to 1.0$\times$10$^{-5}$ mbar, to measure the changes of the crystalline structure, the Fe valence, and the temperature dependence of the resistivity. Our objective is hereby to find the optimal oxygen pressure for the growth of fully stoichiometric single crystalline Fe$_{3}$O$_{4}$ films.

\begin{figure*}
    \centering
    \includegraphics[width=15cm]{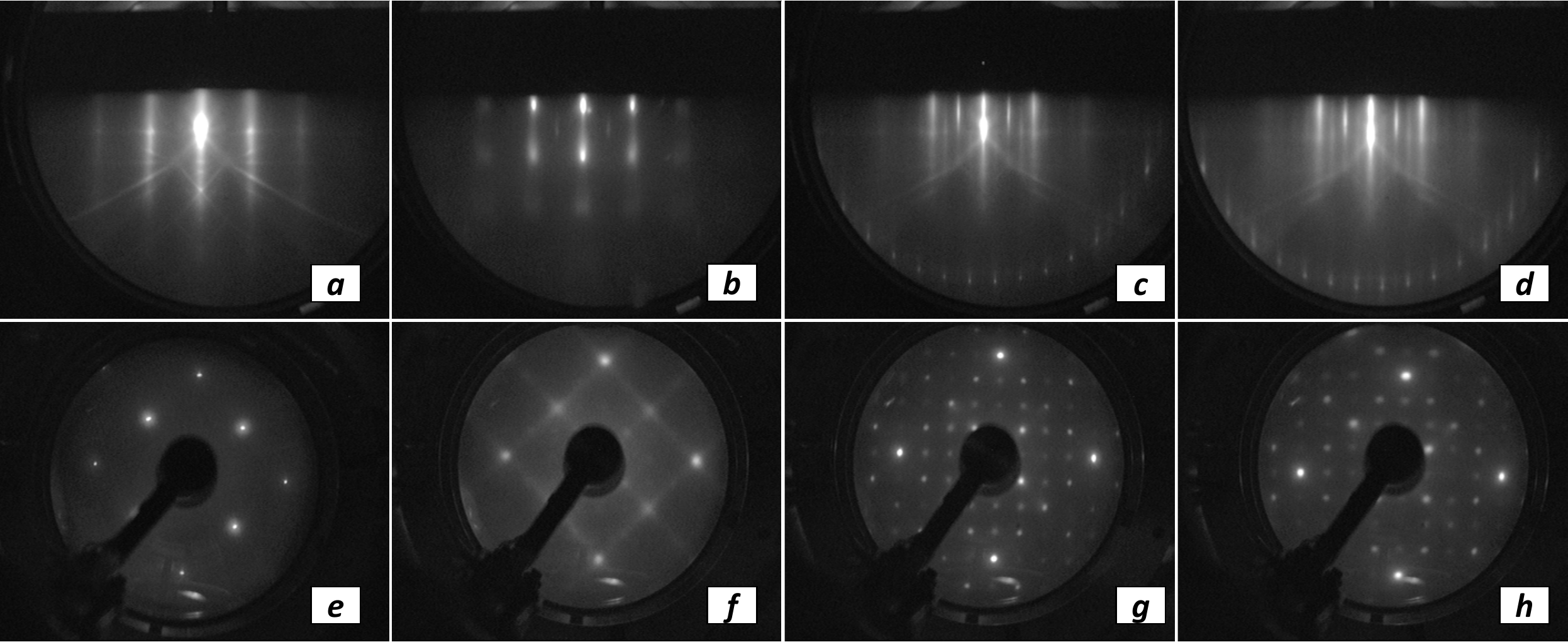}
 \caption{RHEED and LEED electron diffraction patterns of the following: the clean MgO~(001) substrate (a) and (e); 40~nm Fe$_3$O$_4$ films grown at $\Phi$$_{\text{Fe}}$ = 1~\textrm{\AA}/min, T$_{\text{substrate}}$ = $250\,^{\circ}\mathrm{C}$ and P$_{O_2}$ = 5$\times$10$^{-8}$ mbar (b) and (f); 1.0$\times$10$^{-6}$ mbar (c) and (g); 1.0$\times$10$^{-5}$ mbar (d) and (h).}
    \label{Fig1}
\end{figure*}

Figure~1 shows a selection of the RHEED and LEED patterns of the films that we have prepared, namely the clean MgO~(001) substrate [Fig.~\ref{Fig1}(a) and (e)], and Fe$_{3}$O$_{4}$ thin films grown at 5$\times$10$^{-8}$ mbar (b) and (f), 1.0$\times$10$^{-6}$ mbar (c) and (g), and 1.0$\times$10$^{-5}$ mbar (d) and (h) oxygen pressure, respectively. RHEED patterns were taken at 20~keV electron energy, with the beam aligned parallel to the [100] direction of the substrate. The LEED patterns were recorded at an electron energy of 88 eV. The patterns and the quality thereof changes clearly with the oxygen pressure.

For the 1.0$\times$10$^{-6}$ mbar oxygen pressure we observed the characteristic surface structure of bulk Fe$_{3}$O$_{4}$ (001). Sharp RHEED streaks and the presence of Kikuchi lines, as well the intense and sharp LEED spots [Fig.~1(c) and (g)] indicate a flat and well ordered (001) crystalline surface structure of the 40~nm of Fe$_{3}$O$_{4}$ film. The lattice parameter of Fe$_{3}$O$_{4}$ (8.39~\textrm{\AA}) is nearly twice that of MgO (4.21~\textrm{\AA}), resulting in a very small lattice mismatch of 0.3$\%$. Because the growth is fully coherent, with the in-plane dimensions of the spinel unit cell of Fe$_{3}$O$_{4}$ being twice those of rocksalt unit cell of MgO, one expects a new sets of diffraction rods (spots) occurring with half spacing of the substrate. The RHEED and LEED patterns indeed show the occurrence of the half order diffraction rods (spots) in the zeroth Laue zone. The signature of the ($\sqrt{2}\times\sqrt{2}$)R45$^{\circ}$ surface reconstruction, which is also characteristic for single crystal magnetite, is a new set of diffraction rods which are positioned exactly in between the half-order rods (see Fig.~1(c)). We observed a clear and sharp reconstruction pattern both in RHEED and LEED electron diffraction for the film grown at 1.0$\times$10$^{-6}$ mbar oxygen pressure.

For the low 5$\times$10$^{-8}$ mbar oxygen pressure, the RHEED pattern shows transmission like characteristics. The half order diffraction rods are also not clearly visible. The LEED spots become broadened and less intense, indicating an appreciable disorder. Both RHEED and LEED do not show the characteristic crystalline structure of pure Fe$_{3}$O$_{4}$. It is conceivable that the film may also contain FeO and even Fe metal. Photoemission measurements, which will be discussed later, confirm these assumptions.

On the other hand, at the high 1.0$\times$10$^{-5}$ mbar oxygen pressure, the surface reconstruction of Fe$_{3}$O$_{4}$ is still visible in RHEED and LEED patterns, but the electron diffraction patterns are slightly broadened, indicating some increasing disorder in this film. A partial transformation of Fe$_{3}$O$_{4}$ to ${\gamma}$-Fe$_{2}$O$_{3}$, which has the same inverse spinel structure as magnetite, may have even taken place.

\begin{figure}
    \centering
    \includegraphics[width=8cm]{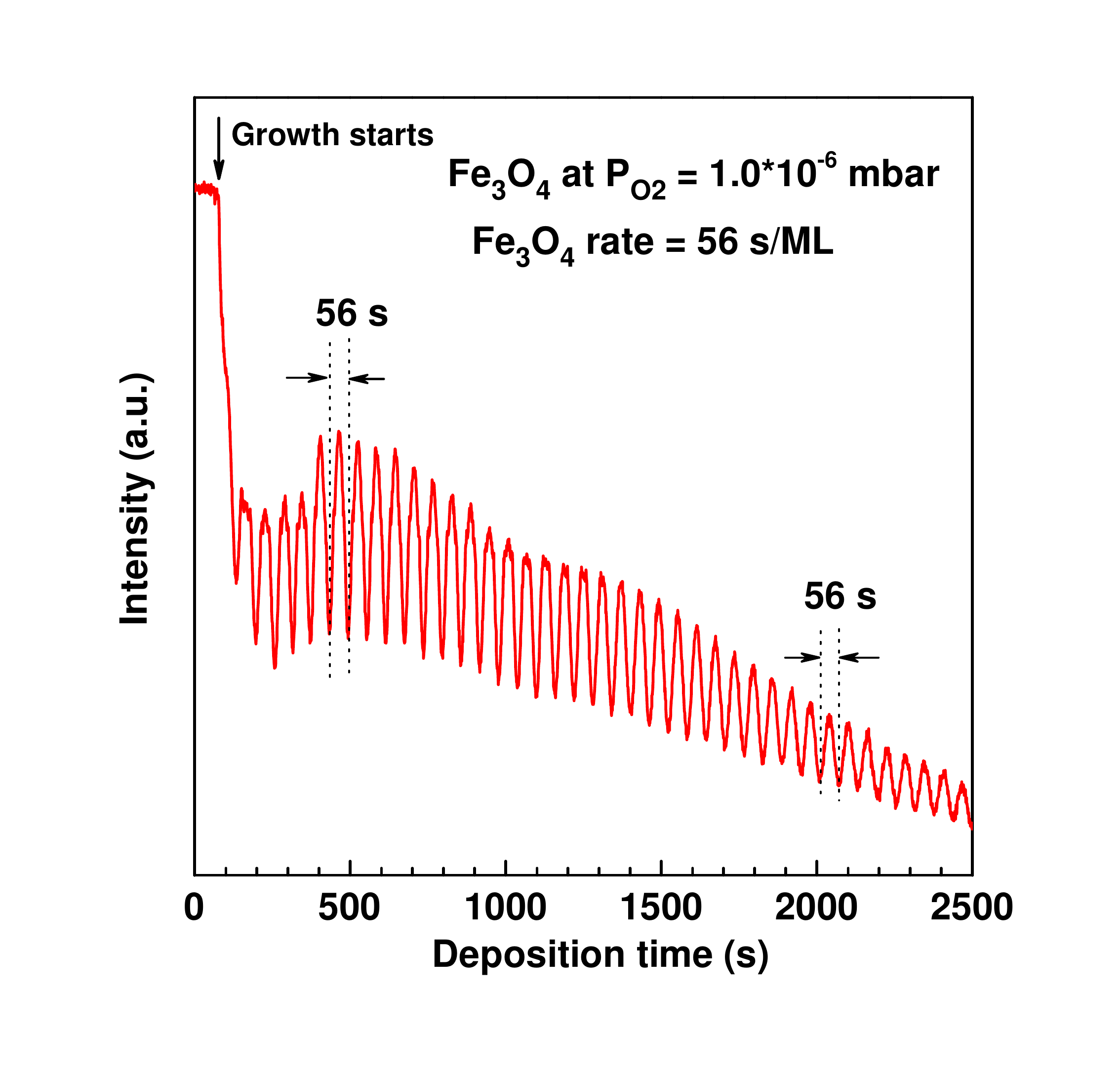}
 \caption{(Color online) RHEED intensity oscillations of the specularly reflected electron beam recorded during deposition of Fe$_3$O$_4$ on MgO~(001) substrate at $\Phi$$_{\text{Fe}}$ = 1~\textrm{\AA}/min, P$_{O_2}$ = 1.0$\times$10$^{-6}$ mbar, and T$_{\text{substrate}}$ = $250\,^{\circ}\mathrm{C}$.}
    \label{Fig2}
\end{figure}

We have also recorded the time development of the crystalline structure during the growth of the films. As an example, we show in Fig.~2 the intensity of the specularly reflected RHEED beam for a Fe$_{3}$O$_{4}$ film grown under 1.0$\times$10$^{-6}$ mbar oxygen pressure. We can clearly observe pronounced intensity oscillations, which are indicative of a two-dimensional  layer-by-layer growth. The time period of the oscillations is 56~s. This period corresponds to the time needed to grow 1~ML of Fe$_{3}$O$_{4}$ and allows for a precise thickness determination.

\begin{figure}
    \centering
    \includegraphics[width=8cm]{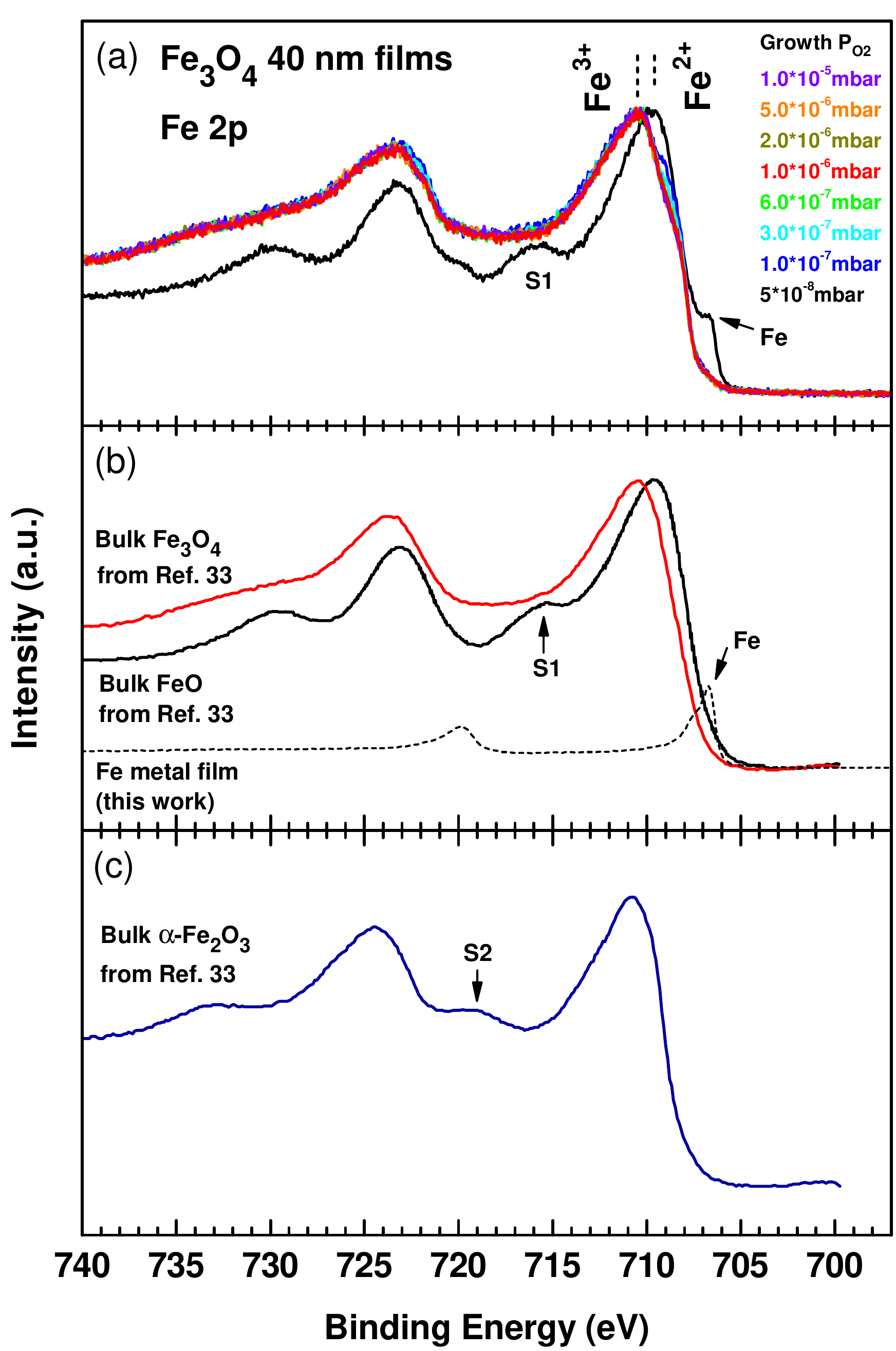}
 \caption{(Color online) XPS Fe~${2p}$ core-level spectra of the following: (a) 40~nm epitaxial Fe$_3$O$_4$ films grown on MgO~(001) by keeping the Fe flux constant and varying the oxygen pressure from 5.0$\times$10$^{-8}$ to 1.0$\times$10$^{-5}$ mbar; (b) bulk Fe$_3$O$_4$ and bulk FeO, reproduced from Ref.~\onlinecite{Moussy2013}, and Fe metal film; (c) bulk $\alpha$-Fe$_2$O$_3$, reproduced from Ref.~\onlinecite{Moussy2013}}
    \label{Fig3}
\end{figure}

In order to clarify the chemical state of the iron oxide, the Fe $2p$ core level XPS spectra were recorded for the Fe$_{3}$O$_{4}$ films grown under varying oxygen pressures, i.e., from 5$\times$10$^{-8}$ to 1.0$\times$10$^{-5}$ mbar, while the Fe flux and the thickness are kept constant. The results are plotted in Fig.~3(a). We also display in Fig.~3(b) the XPS spectra of bulk Fe$_3$O$_4$ \cite{Moussy2013}, bulk FeO \cite{Moussy2013}, and Fe metal film, and in Fig.~3(c) of bulk $\alpha$-Fe$_2$O$_3$\cite{Moussy2013} as reference. FeO has a clear satellite feature at 715.5 eV, marked as S1 in Fig.~3(b), while Fe$_2$O$_3$ shows a satellite feature at 719.1 eV, marked as S2 in Fig.~3(c). Such satellite structures are frequently used as fingerprints to identify the other iron oxide phases \cite{Chambers1997,Gao97,Fuji1999,Moussy2013}. One can clearly see that with the exception of the sample grown at the low 5$\times$10$^{-8}$ mbar pressure, the Fe $2p$ spectrum for all the other Fe$_{3}$O$_{4}$ films show no signs for satellite structures S1 and S2.

The main peaks for Fe$_{3}$O$_{4}$ are relatively broad since they are given by the three different Fe contributions in Fe$_3$O$_4$, namely the tetrahedral Fe$^{3+}$ and the octahedral Fe$^{2+}$ and Fe$^{3+}$ sites. The characteristic energies for the Fe$^{2+}$ and Fe$^{3+}$ are marked in Fig.~3. Going from high to low pressures: the spectra of the films prepared using 1.0$\times$10$^{-5}$ to 6.0$\times$10$^{-7}$ mbar all look identical and have very similar line shapes as the ones reported for magnetite in the literature \cite{Fuji1999,Chambers1997,Moussy2013}; see the top curve in Fig.~3b. Comparing with the spectrum of Fe$_2$O$_3$ in Fig.~3(c) we also learn that a molecular oxygen pressure up to 1$\times$10$^{-5}$ mbar is still not enough to form the Fe$_2$O$_3$ phase. Subtle changes at 709 eV energy can be observed for films grown at 3.0$\times$10$^{-7}$ and 1.0$\times$10$^{-7}$ oxygen pressures, suggesting an increase of the Fe$^{2+}$ content associated with the presence of oxygen defects. The Fe $2p$ XPS spectra of the film grown at the lowest oxygen pressure of the series, 5$\times$10$^{-8}$ mbar, looks very much different and the feature at 707 eV (indicated by the "Fe" label) can be taken as an indication that metallic Fe clusters are present since this 707 eV feature is the main peak in the XPS of Fe metal film as plotted in Fig.~3b.

\begin{figure}
    \centering
    \includegraphics[width=8cm]{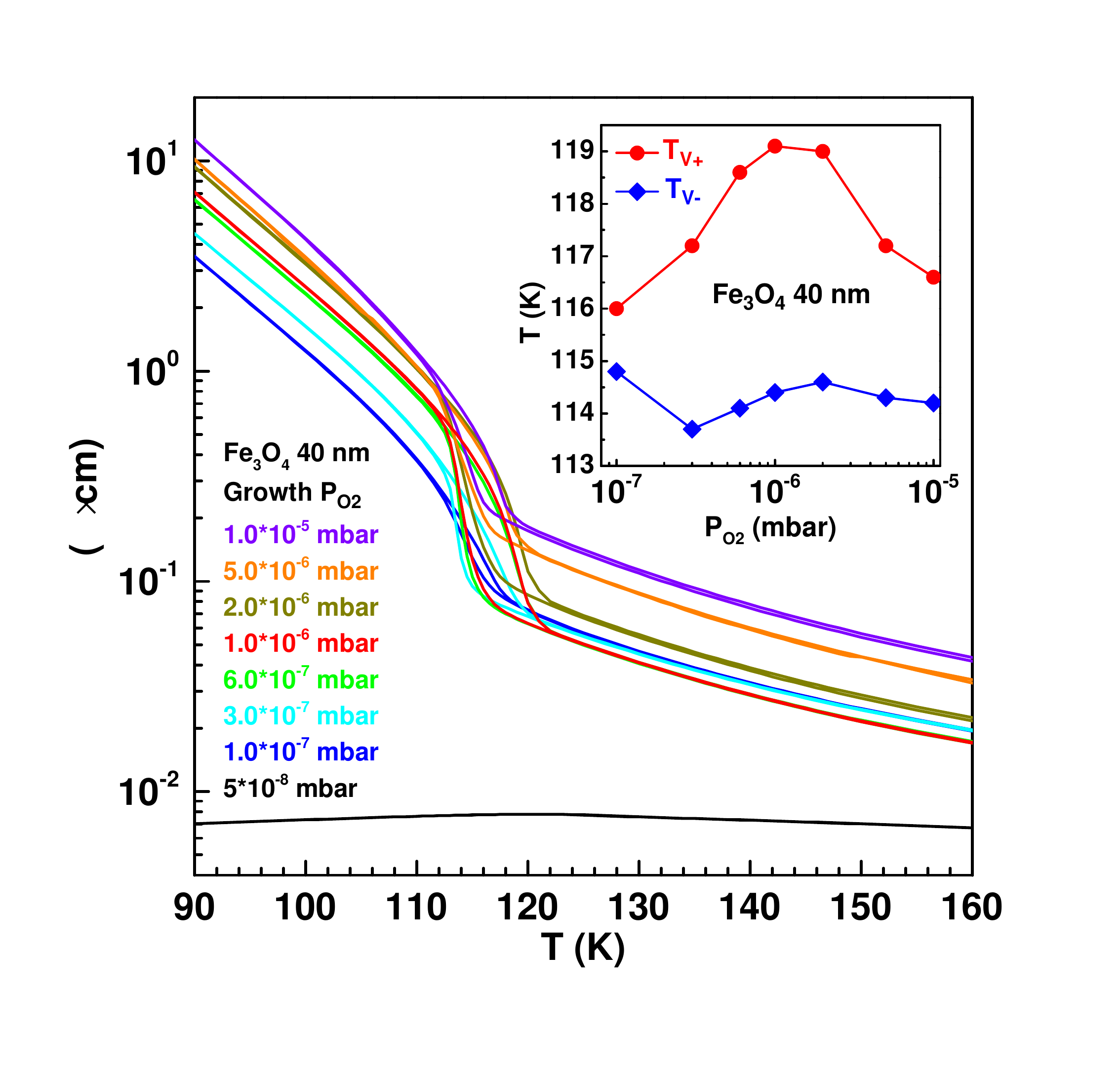}
 \caption{(Color online) Resistivity as a function of temperature for 40~nm Fe$_3$O$_4$ films grown under various O$_2$ partial pressures P$_{O_2}$. Inset: the Verwey transition temperatures (\textit{T}$_{V+}$ and \textit{T}$_{V-}$) (see text for their definition) are plotted as a function of P$_{O_2}$.}
    \label{Fig4}
\end{figure}

In order to determine the optimum oxygen pressure for the growth of fully stoichiometric Fe$_{3}$O$_{4}$ film, we performed temperature dependent resistivity measurements on this set of films. Figure~\ref{Fig4} shows that all films except the one grown under 5$\times$10$^{-8}$ mbar oxygen pressure exhibit a transition in the resistivity with a clear hysteresis. The Verwey transition is clearly first order in these films. The observation that the sample grown at P$_{O_2}$ = 5$\times$10$^{-8}$ mbar does not show any metal-insulator transition is consistent with the results of the electron diffraction (RHEED and LEED) and XPS measurements, which all indicate the presence of a different phase.

We now define the Verwey transition temperature \textit{T}$_{V-}$ and \textit{T}$_{V+}$ as the temperature of the maximum slope of \textit{$\rho$}~(\textit{T}) curve for the cooling down and warming up temperature branch, respectively. These transition temperatures are plotted as a function of the oxygen pressure P$_{O_2}$ of the films as inset in Fig.~\ref{Fig4}. We can observe that \textit{T}$_{V+}$ varies between 116 K and 119 K, and that \textit{T}$_{V-}$ ranges between 113.5 K and 115 K. The highest transition temperature is reached for the sample grown at 1.0$\times$10$^{-6}$ mbar oxygen pressure, with a \textit{T}$_{V+}$ of 119 K. We therefore choose this as the optimal oxygen pressure for the growth, also because the RHEED and LEED patterns are the best in terms of sharpness, indicating a good crystalline surface structure.

\begin{figure}
    \centering
    \includegraphics[width=8cm]{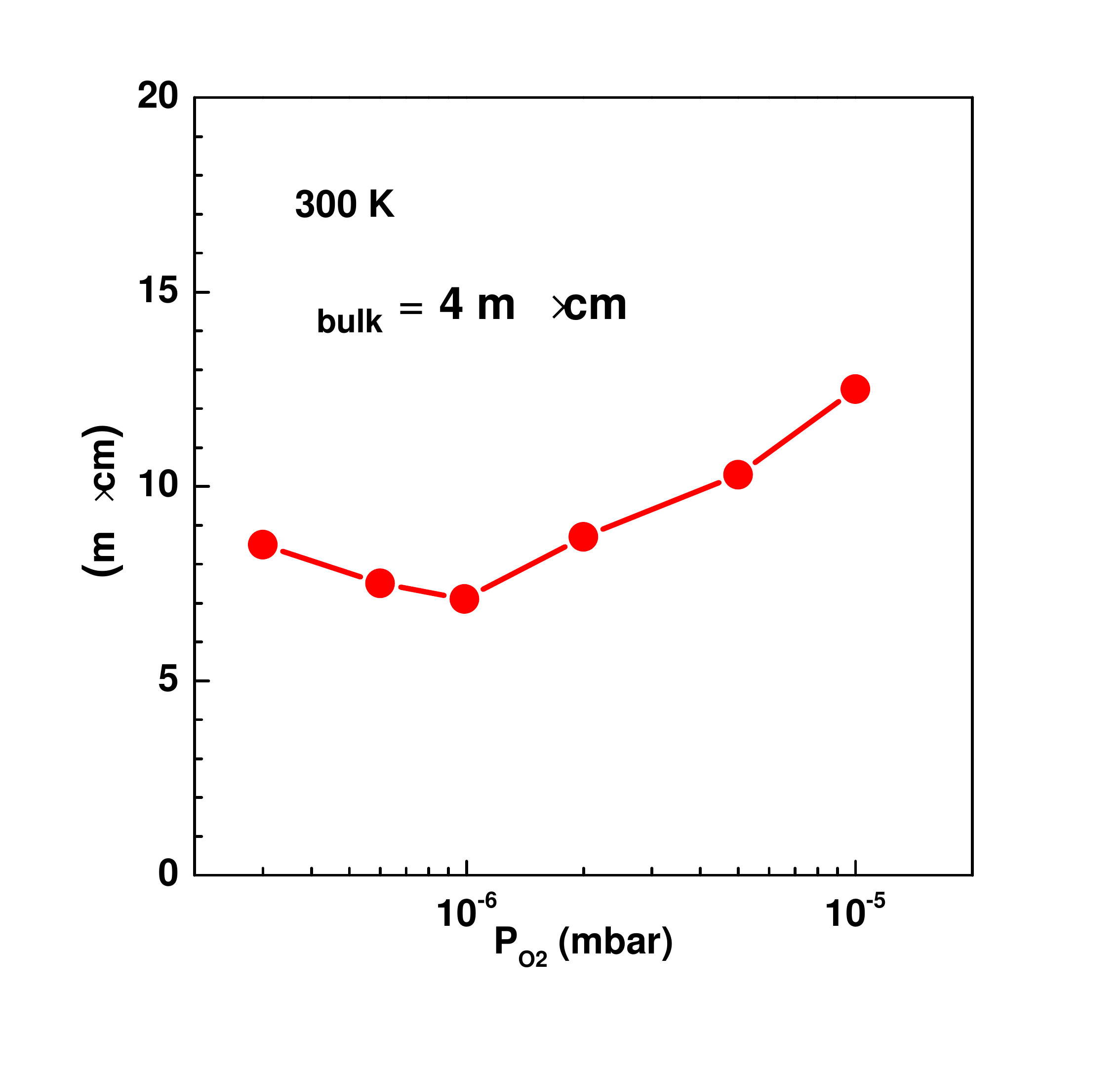}
 \caption{(Color online) Room temperature values of the resistivity of 40~nm Fe$_3$O$_4$ films as a function of P$_{O_2}$.}
    \label{Fig5}
\end{figure}

We calculated the absolute value of the room temperature resistivity of our thin films and plotted the results as a function of oxygen pressure in Fig.~\ref{Fig5}. Interestingly, we found a minimum of the resistivity for P$_{O_2}$ = 1.0$\times$10$^{-6}$ mbar \cite{Kiyomura1999}. The magnetite film prepared at the optimum oxygen pressure has a room temperature resistivity of about 7 m$\Omega$-cm, compared to 4 m$\Omega$-cm measured in bulk Fe$_{3}$O$_{4}$ single crystal \cite{Miles57}. The higher resistivity values shown by the Fe$_{3}$O$_{4}$ films compared to the bulk may be attributed to the presence of antiphase boundaries (APBs). It was reported that APBs are likely to form in  Fe$_3$O$_4$ thin films grown on MgO substrates \cite{Eerenstein02,Kumar06} because the larger unit cell of Fe$_3$O$_4$ (\textit{Fd}$\overline{3}$m) in comparison to that of MgO (\textit{Fm}3m) makes that nucleation sites equivalent on the MgO are not equivalent for the Fe$_3$O$_4$. Domains of Fe$_3$O$_4$ are therefore formed with APBs between them, and as the magnetic coupling over a large fraction of these boundaries is antiferromagnetic,\cite{Eerensteinprl02} these APBs act as scattering centers that hinder electron transport \cite{Eerenstein02,Kumar06,Ramos06,Sofin11}.

\section{Thickness dependence}

\begin{figure}
    \centering
    \includegraphics[width=8cm]{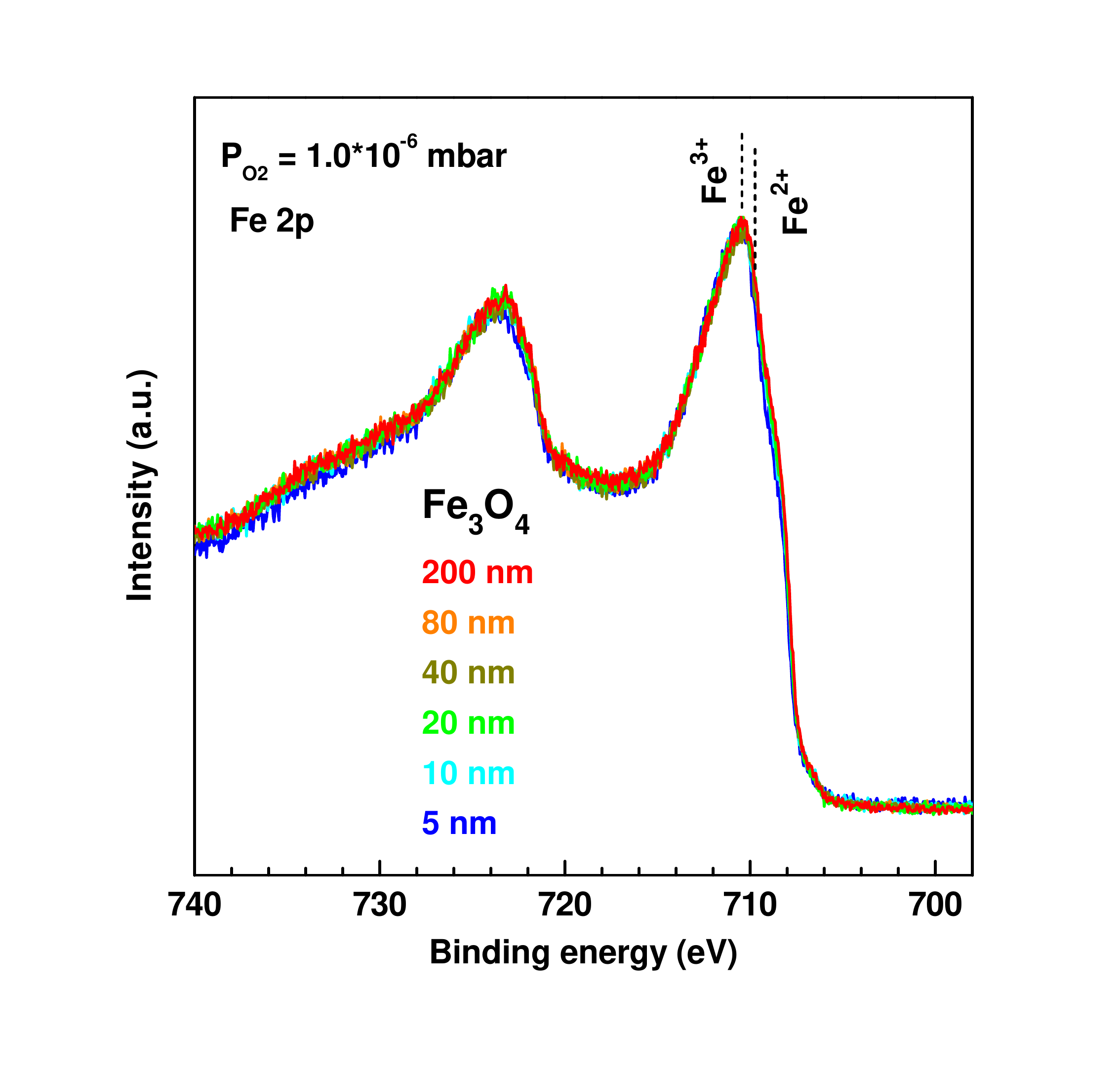}
 \caption{(Color online) Fe~${2p}$ XPS spectra of 5-, 10-, 20-, 40-, 80-, and 200-nm-thick epitaxial Fe$_3$O$_4$ films on MgO~(001).}
    \label{Fig6}
\end{figure}

Having established the optimum oxygen pressure for growing fully stoichiometric Fe$_3$O$_4$ films, we proceed with the investigation on the effect of the film thickness on the Verwey transition. A new series of Fe$_3$O$_4$ thin films with different thicknesses (5, 10, 20, 40, 80, and 200~nm) was grown on MgO~(001) substrates, all at P$_{O_2}$ = 1$\times$10$^{-6}$ mbar. The Fe flux is set at 1~\textrm{\AA} /min and the substrate temperature at $250\,^{\circ}\mathrm{C}$. We verified the crystalline quality using RHEED and LEED. Sharp RHEED streaks of the surface reconstruction and the presence of Kikuchi lines indicate that the surface is very smooth, even after deposition of 200~nm Fe$_{3}$O$_{4}$.

Figure~\ref{Fig6} presents the Fe $2p$ core level XPS spectra of the Fe$_3$O$_4$ films as a function of thickness. Overall, the XPS spectra show the characteristics of stoichiometric magnetite. There are no indications for the presence of other iron oxide phases. Only in the very thin film (5~nm) a slightly smaller Fe$^{2+}$ contribution to the Fe~$2p$ main peaks can be noticed, otherwise no significant changes are observed with increasing the thickness from 10~nm to 200~nm.

\begin{figure}
    \centering
    \includegraphics[width=8cm]{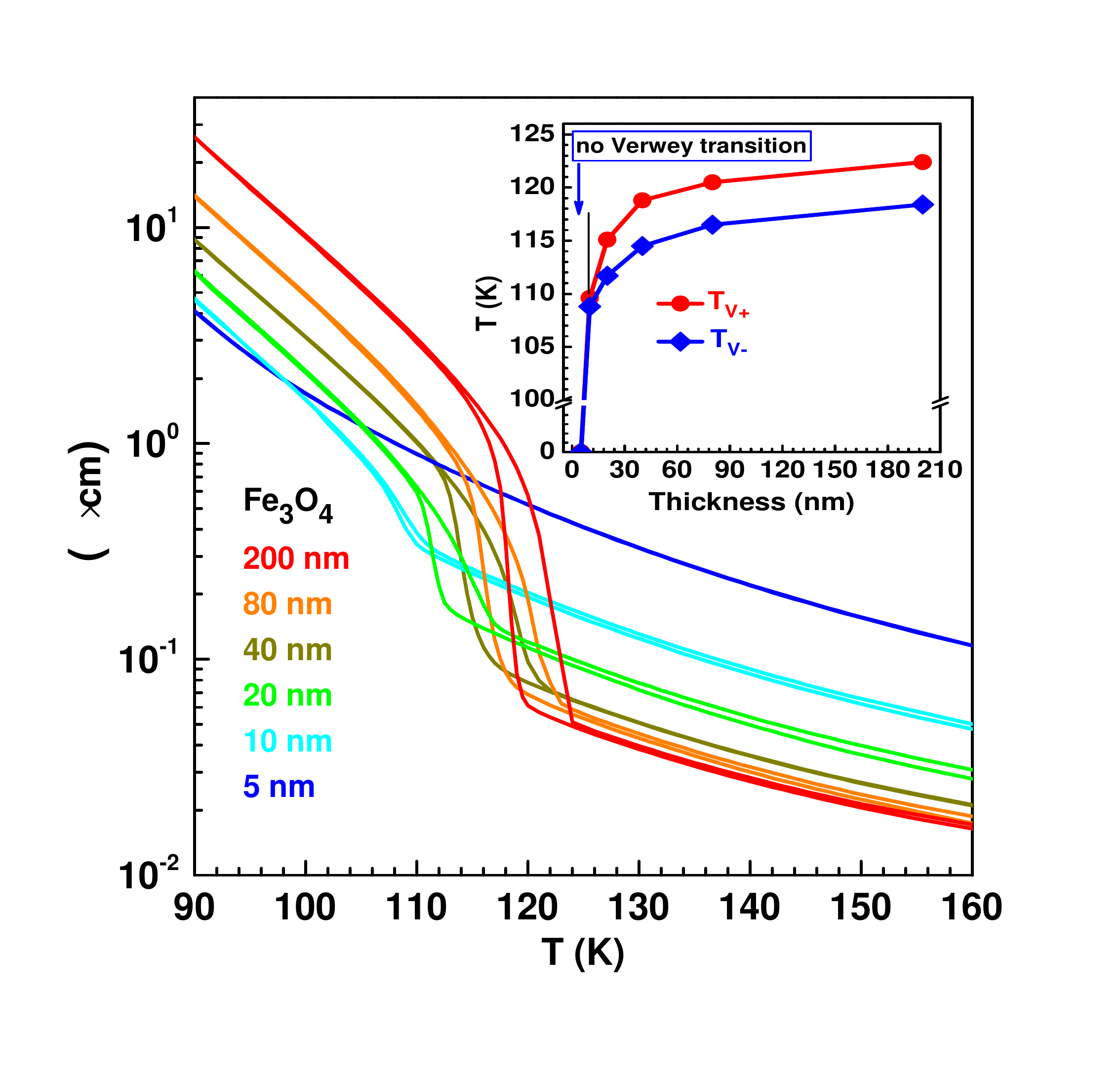}
 \caption{(Color online) Resistivity as a function of temperature for Fe$_3$O$_4$ films with different thicknesses. Inset: the Verwey transition temperature (\textit{T}$_{V+}$ and \textit{T}$_{V-}$) as a function of film thickness.}
    \label{Fig7}
\end{figure}

Figure~\ref{Fig7} displays the resistivity as a function of temperature for Fe$_3$O$_4$ films with different thicknesses grown on the MgO substrates. The 5~nm thick Fe$_3$O$_4$ film does not show a transition, while the 10 nm and thicker films all reveal a transition with hysteresis, establishing again that the Verwey transition in these films is first order. The transition becomes more pronounced and the transition temperature gets higher with increasing the thickness. The change in resistivity is the largest for the 200~nm Fe$_3$O$_4$ film (the thickest film grown in this study), with also the highest transition temperature (\textit{T}$_{V+}$) of about 122~K, similar to the value reported for much thicker films (660~nm) \cite{Li98} and close to the value reported for bulk samples \cite{Garcia04,Miles57}. 

\textcolor[rgb]{0.98,0.00,0.00}{
\begin{figure}
    \centering
    \includegraphics[width=8cm]{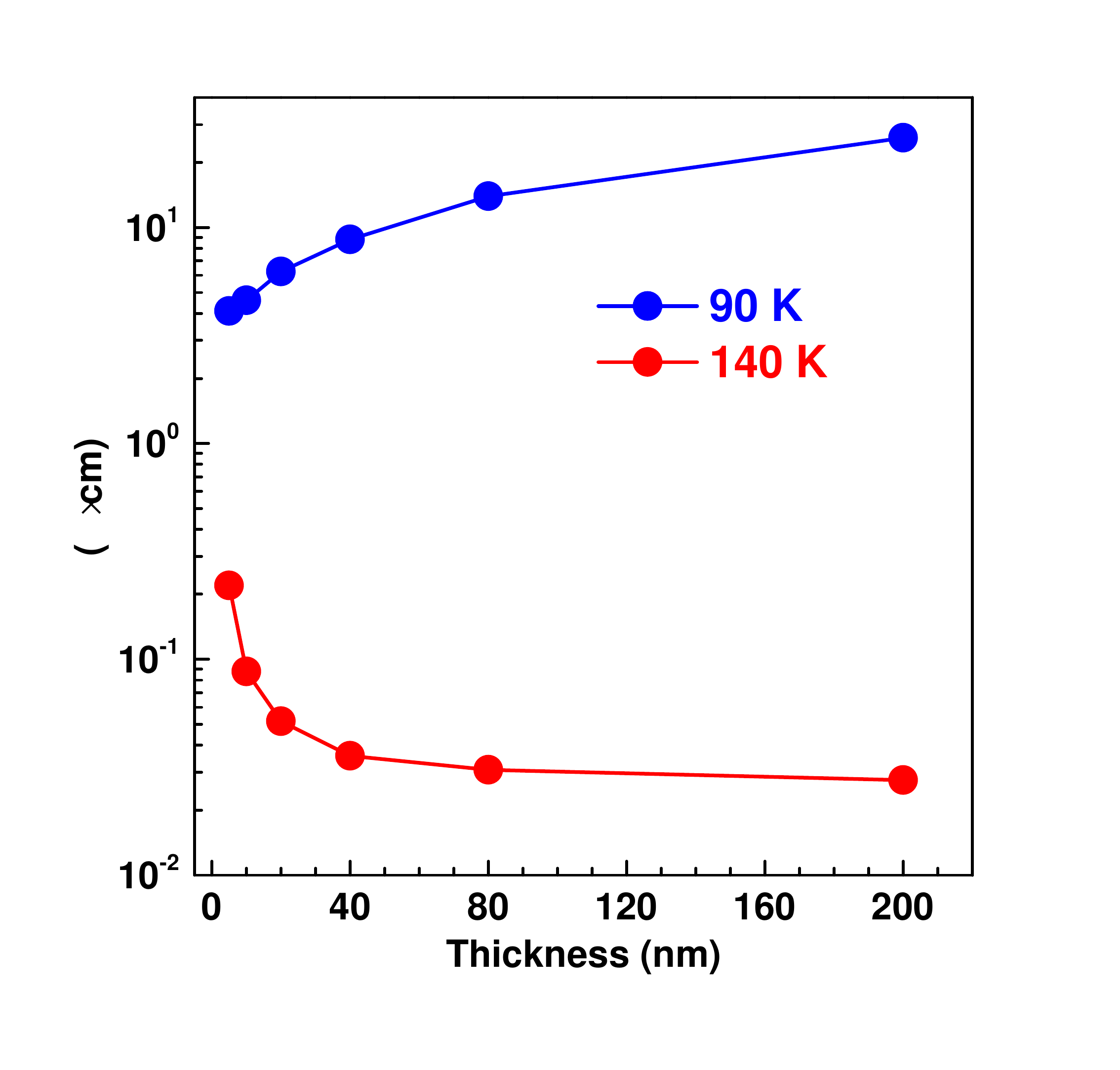}
 \caption{(Color online) Resistivity of the Fe$_3$O$_4$ films at 90 K (blue) and 140 K (red) as a function of thickness.}
    \label{Fig8}
\end{figure}
}

It is interesting to note that the resistivity above the transition temperature is the highest for the thinnest films, but that the situation is opposite below the transition temperature, namely, the highest resistivity is for the thickest films. See Fig.~\ref{Fig8} which displays the resistivity at 90 K (below the transition) and 140 K (above the transition) as a function of film thickness. The influence of scattering due to imperfections in the vicinity of the surface and/or interface with the substrate will naturally be larger for the thinner films relative to the thicker ones, so it is to be expected that the thinner films will have higher resistivities. This is the case for temperatures above the transition. However, the situation below the transition requires a different explanation. Apparently the band gap or conductivity gap that can be opened in the low temperature phase is smaller in the thinner films. In fact, the data indicates that such a gap opening does not occur at all in the 5 nm film so that it becomes more conducting than the thicker films in the low temperature phase.

The Verwey transition temperatures \textit{T}$_{V+}$ and \textit{T}$_{V-}$ as a function of thickness are plotted in the inset of Fig.~\ref{Fig7}. The transition temperature increases rapidly with the initial increase in the film thickness and then gradually approaches a value close to the bulk. Furthermore, the high temperature values of resistivity gradually decrease with increasing the thickness. We would like to note that the transitions shown in Fig.~\ref{Fig7} are generally sharper than those reported in earlier Fe$_3$O$_4$ thin film studies; see, for comparison, Refs. [\onlinecite{Eerenstein02,Kumar06,Li98,Ramos06,Sofin11}].

The reduction of \textit{T}$_{V}$ and the broadening of the Verwey transition for thinner films was attributed previously to the residual strain \cite{Li98} and the suppression of the orthorhombic deformation which takes place at the Verwey transition \cite{VanDeVeerdonk}. Eerenstein \textit{et al} \cite{Eerenstein02} have shown that the increase of the resistivity with decreasing film thickness can be related to a strong increase in APBs density, thus a significant decrease in domain size. By using transmission electron microscopy (TEM) they found that the domain size decreases from 40~nm for 100-nm-thick films, to 5~nm for 3-nm-thick films. It was suggested that the absence of the Verwey transition in films thinner than 25~nm may be also related to very small domain size \cite{Eerenstein02}. In contrast to the results reported by Eerenstein \textit{et al} \cite{Eerenstein02}, we still observed a clear Verwey transition from a 10~nm thick film.

\begin{figure}
    \centering
    \includegraphics[width=8cm]{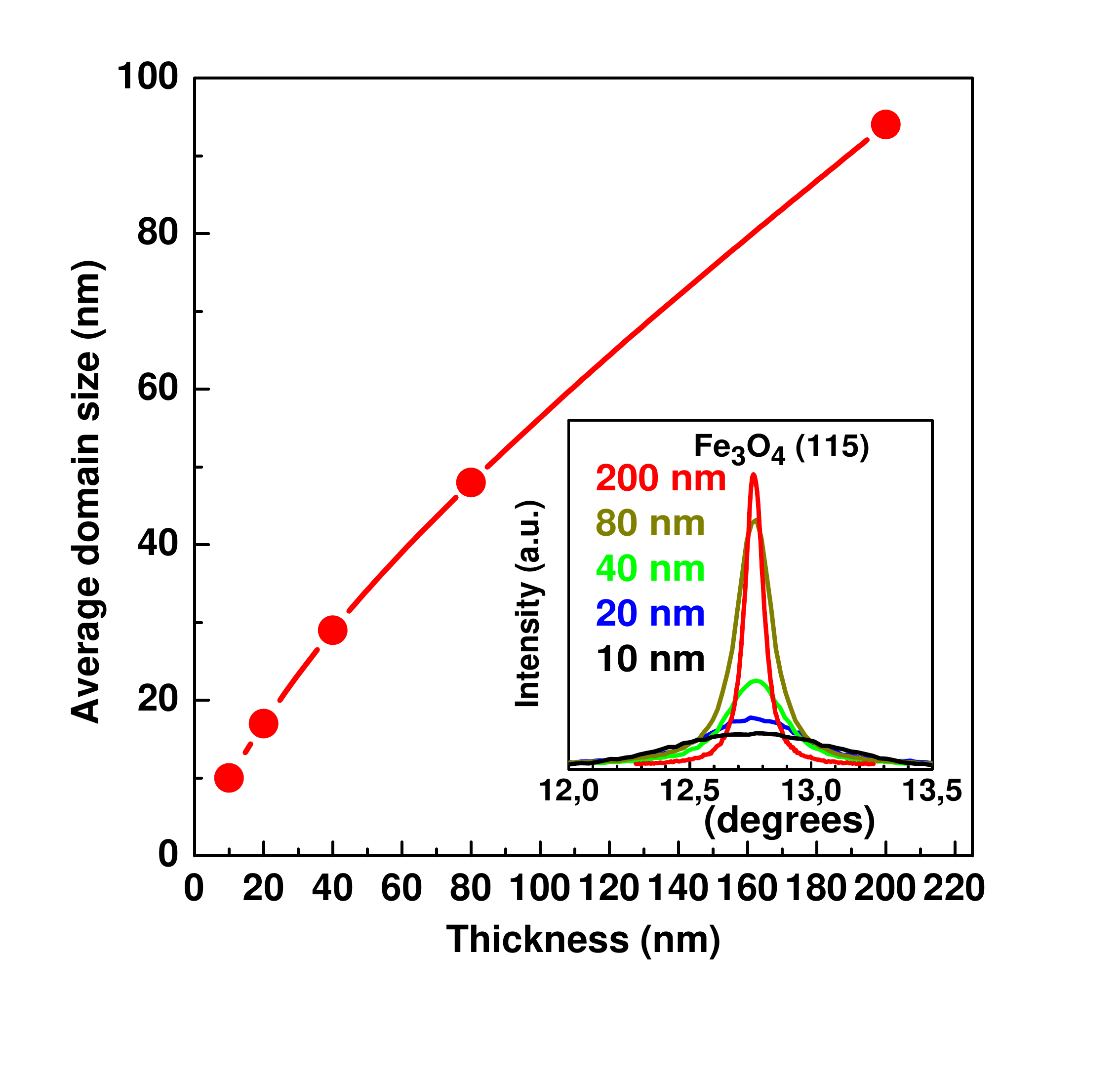}
 \caption{(Color online) Structural average domain size obtained from x-ray diffraction versus thickness. Inset: high resolution rocking curves of the (115) peak of 10-, 20-, 40-, 80-, and 200-nm-thick Fe$_3$O$_4$ films.}
    \label{Fig9}
\end{figure}

\begin{figure}
    \centering
    \includegraphics[width=8cm]{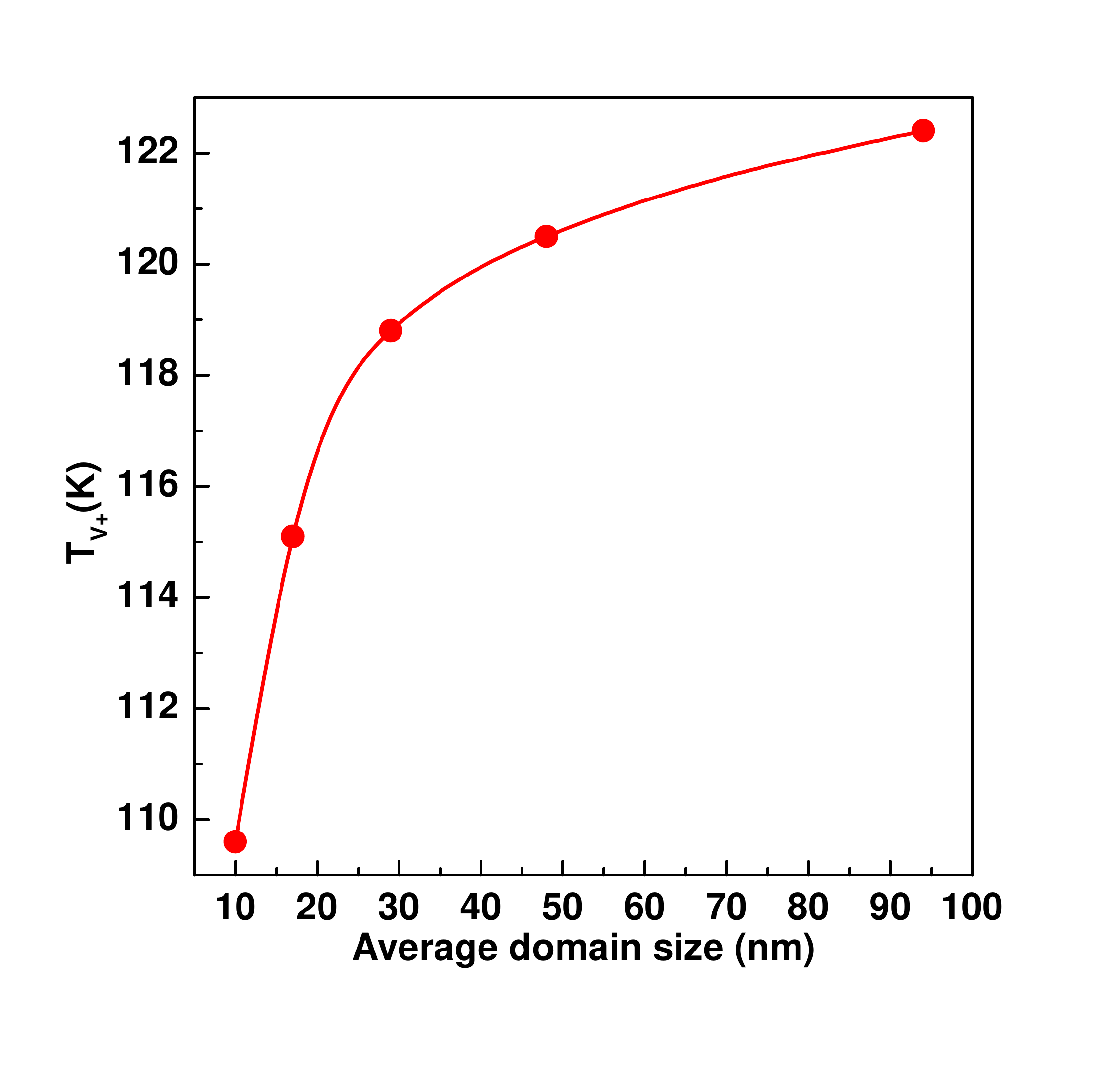}
 \caption{(Color online) \textit{T}$_{V+}$ as a function of average domain size of the 10-, 20-, 40-, 80-, and 200-nm-thick Fe$_3$O$_4$ films.}
    \label{Fig10}
\end{figure}

To better understand the effect of the thickness on the Verwey transition in Fe$_3$O$_4$ thin films we looked at the microstructure of our films by high-resolution x-ray diffraction (HR-XRD) and determined the relative changes of the average domain size with thickness. XRD rocking-curve measurements of the in-plane reflection (115) of the 10-, 20-, 40-, 80-, and 200-nm-thick Fe$_3$O$_4$ films grown on MgO are displayed as an inset in Fig.~\ref{Fig9}. A sharp decrease of the full width at half maximum (FWHM) with increasing the thickness is observed. By using the simple Scherrer formula \cite{Scherrer,Boulch2001,Patterson1939} we calculated the average domain size (ADS) taking into account the instrumental broadening. Here we note that fitting the profiles with a Voigt function yields an appreciable Gaussian contribution (beyond that of the instrumental broadening) which indicates the presence of mosaicity in addition to the Lorentzian contribution which represents the average domain size \cite{Patterson1939}. Nevertheless, for the purpose of obtaining a crude characterization, we take the presence of mosaicity also as a sign for a decreased domain size. The ADS as a function of thickness is plotted in Fig.~\ref{Fig9}. It increases from 10~nm for the 10-nm-thick film to 94~nm for the 200-nm-thick film, a trend that agrees well with the results of other groups \cite{Eerenstein02,Ramos06,Eerenstein03}. The fact that the domain size plays an important role and greatly influences the Verwey transition of Fe$_3$O$_4$ thin films is nicely illustrated in Fig.~\ref{Fig10}, where the \textit{T}$_{V+}$ is plotted as a function of the ADS. This result strongly suggests that the larger the ADS of Fe$_3$O$_4$ films the higher the transition temperature \textit{T}$_{V}$ and the larger the conductivity gap that can be opened in the low temperature phase.

\section{Strain and microstructure effects}

\begin{figure}
    \centering
    \includegraphics[width=8cm]{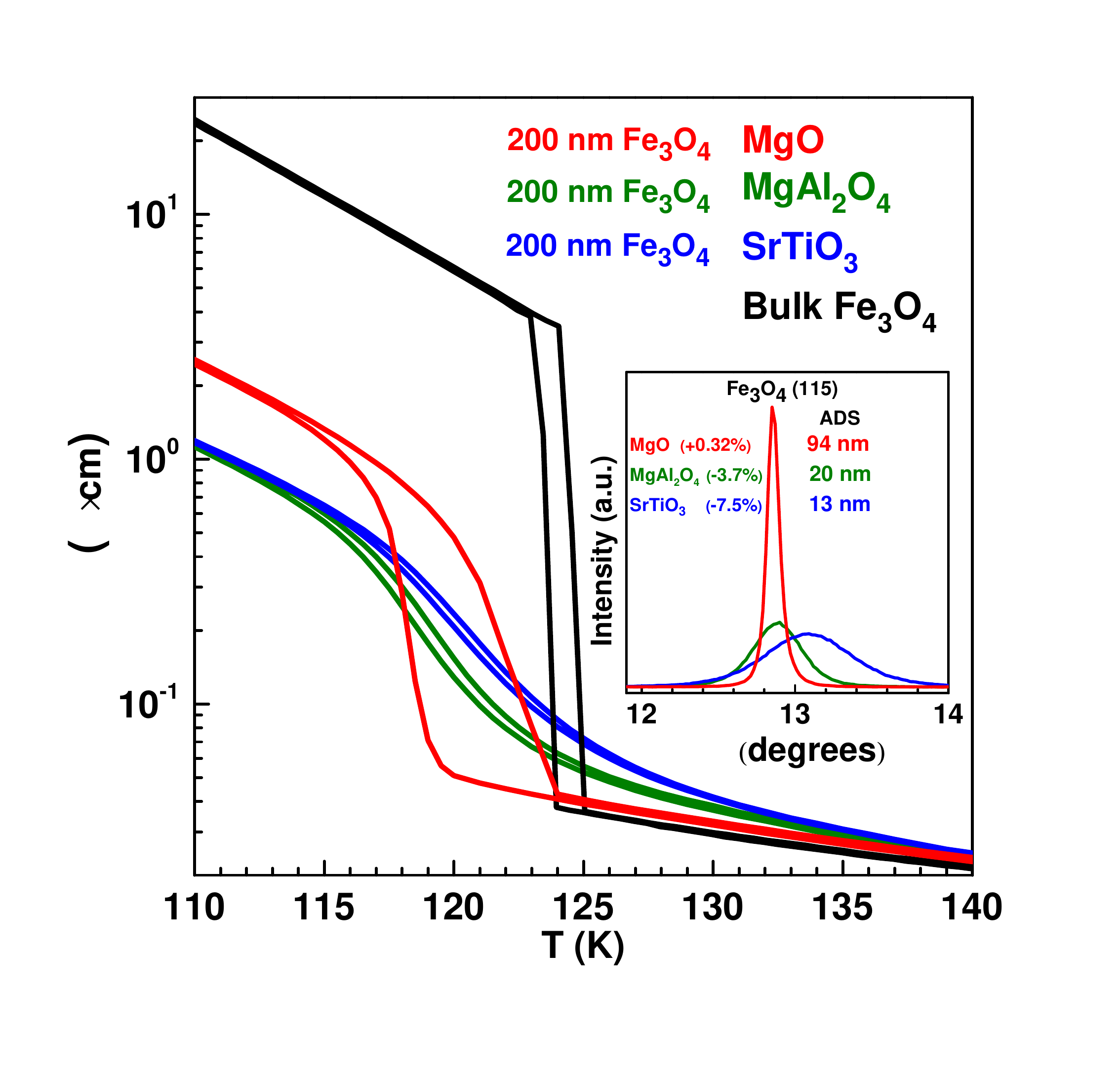}
 \caption{(Color online) Resistivity as a function of temperature of 200-nm-thick Fe$_3$O$_4$ films grown on MgO~(001), MAO~(001), and STO~(001) and of the single crystal bulk Fe$_3$O$_4$. Inset: rocking curves of the (115) peak of 200~nm Fe$_3$O$_4$ films grown on MgO~(001), MAO~(001), and STO~(001) substrates.}
    \label{Fig11}
\end{figure}

To investigate whether more factors than only oxygen stoichiometry and thickness have an influence on the Verwey transition, we have also grown films on substrates other than MgO. In particular, we have utilized MgAl$_2$O$_4$ (001) (MAO) and SrTiO$_3$ (001) (STO) substrates. The lattice constant of bulk Fe$_3$O$_4$ is 8.397 \textrm{\AA} \cite{Margulies97}, which is slightly smaller than twice that of MgO with 4.212 \textrm{\AA} \cite{Margulies97}, but larger than that of MAO with 8.08 \textrm{\AA} \cite{Eerensteinjmmm03}, and appreciably larger than twice that of STO with 3.905 \textrm{\AA} \cite{Hamie12}. Figure~\ref{Fig11} compares the temperature dependence of the resistivity of 200 nm Fe$_3$O$_4$ films grown on MgO, MAO, and STO with that of a single crystal Fe$_3$O$_4$ as reference. We note that the XPS spectra of the films grown on MgO, MAO and STO are essentially identical, see Fig.~12, indicating that the substrate has no influence on the chemical composition of the film.

\begin{figure}
    \centering
    \includegraphics[width=8cm]{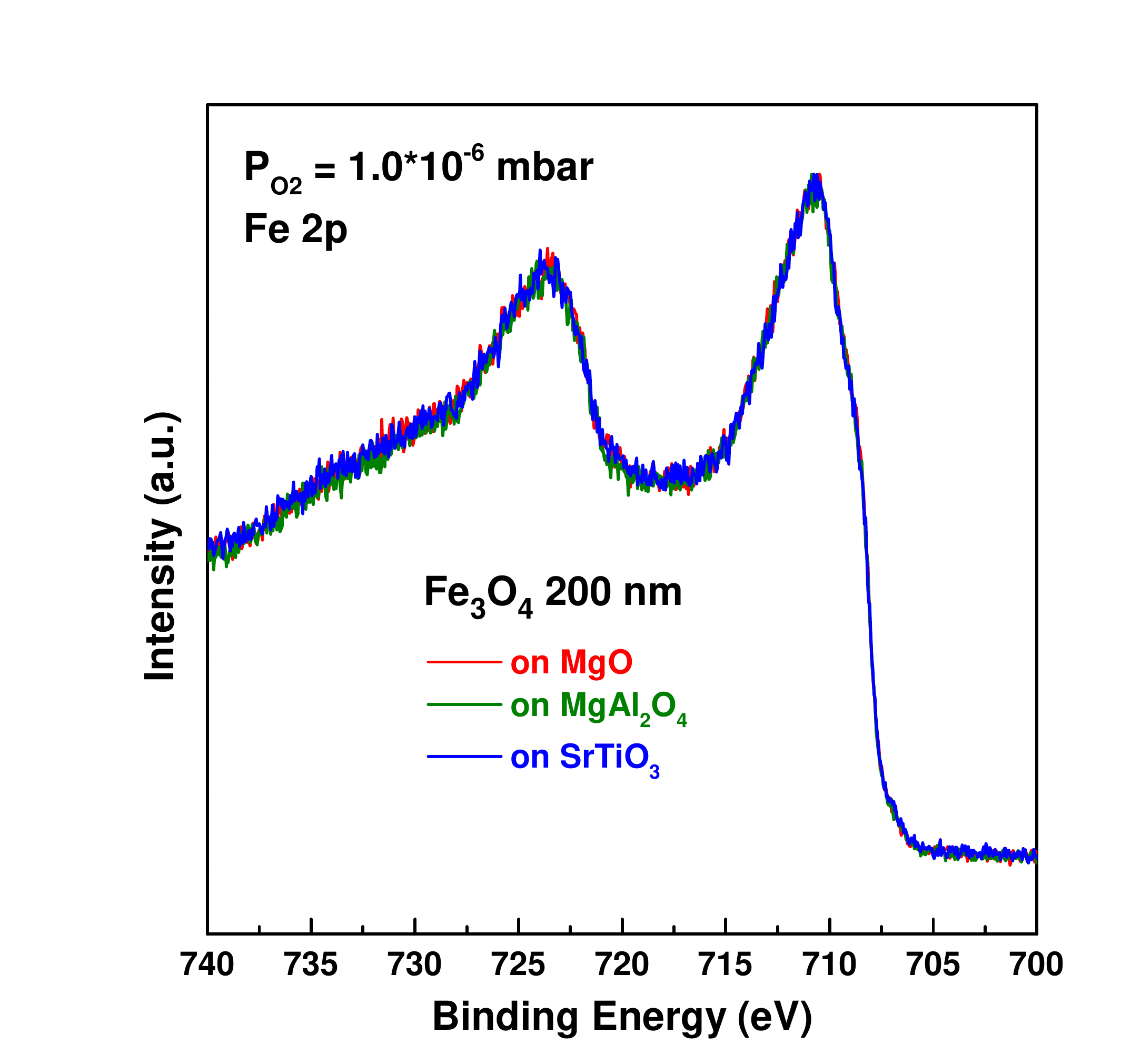}
 \caption{(Color online) Fe $2p$ XPS spectra of 200-nm-thick epitaxial Fe$_3$O$_4$ films on MgO(001),
     MAO(001), and STO(001).}
    \label{Fig12}
\end{figure}

We can observe the Verwey transition in bulk Fe$_3$O$_4$ with a \textit{T}$_{V+}$ of 125 K and a width of about 1 K. The 200 nm Fe$_3$O$_4$ film on MgO has a \textit{T}$_{V+}$ of 122 K with a width of about 3 K. Here the width is defined as the full width at half maximum of the temperature derivative of the resistivity across the transition. The transition for the 200 nm films grown on MAO and STO is by contrast extremely broad with a much less pronounced resistivity change. The width is of order of 10 K, and yet, the temperature curves still show a clear hysteresis, demonstrating that the transition is still first order and not second or higher order \cite{Aragon85,Kakol92,Aragon93,Honig95,Brabers98}. This is an important finding since the observation of a broad first order transition implies that the system is inhomogenous, i.e. it consists of a distribution of crystallites each having its own (sharp) first order transition temperature.

To address the effect of the substrate, we have to consider first of all the lattice mismatch between Fe$_3$O$_4$ and MgO, MAO, and STO, which are +0.32$\%$, -3.77$\%$ and -7.5$\%$, respectively \cite{Margulies97,Eerensteinjmmm03,Hamie12}. While the 200~nm Fe$_3$O$_4$ on MgO remains fully strained up to very high thicknesses \cite{Arora06}, the 200~nm Fe$_3$O$_4$ films on STO and MAO do relax to the bulk structure \cite{Luysberg09} due to the large mismatch. For the latter films one may then expect to see a Verwey transition like that of the bulk material, but the experiment reveals a very different behavior. To explain the observed broadness of the transition we performed high resolution x-ray diffraction measurements. Rocking curves of the (115) reflection of the  Fe$_3$O$_4$ grown on the different substrates clearly show a much broadened  peak for the Fe$_3$O$_4$ films on MAO and STO; see inset in Fig.~\ref{Fig11}. From these profiles we have calculated the average domain size to be about 20~nm for the Fe$_3$O$_4$ on MAO and 13~nm for Fe$_3$O$_4$ on STO, while it is about 94~nm for Fe$_3$O$_4$ on MgO. This result indicates that the strain relaxation of the Fe$_3$O$_4$ films on MAO and STO is accompanied by a breakup into small domains of Fe$_3$O$_4$ with a wide distribution of the domain size.

Another interesting observation that can be made from Fig.~\ref{Fig11} is that the hysteresis is rather small in bulk Fe$_3$O$_4$ as well as in the 200 nm films of Fe$_3$O$_4$ on MAO and STO, i.e., about 1 K, but that the hysteresis is rather large for the 200 nm film on MgO, i.e., about 3 K. Since the hysteresis width does not depend very much on the film thickness, see Fig. 7, and therefore not on the average domain size, see Fig. 9, we deduce that the hysteresis must be connected to growth aspects that make MgO substrates different from MAO and STO. We infer that this is the presence of APBs of Fe$_3$O$_4$ films on MgO associated with equivalent nucleation sites on the MgO separated by nonunit cell vectors of the Fe$_3$O$_4$ lattice. For Fe$_3$O$_4$ films on MAO and STO, on the other hand, we expect the formation of APBs to be much less likely. Indeed, it is quite conceivable that the presence of APBs constitute an important energy barrier for the crystal structure transformation across the Verwey transition.

\section{Conclusion}

To summarize, we have carried out a systematic experimental investigation to identify the parameters that determine the quality of the Verwey transition in epitaxial Fe$_3$O$_4$ thin films. We have maximized the transition temperature by optimizing the oxygen pressure used for the growth of the films and we have found that the substrate-induced microstructure plays a crucial role. The transition temperature, the resistivity jump, and the conductivity gap of fully stoichiometric films greatly depend on the domain size, which increases gradually with increasing film thickness. The broadness of the transition correlates strongly with the width of the domain size distribution. We infer that the width of the hysteresis is influenced strongly by the presence of antiphase boundaries. Films grown on MgO (001) substrates showed the highest and sharpest transitions, while films grown on substrates with large lattice constant mismatch revealed very broad transitions. In all cases in which the Verwey transition is present, the transition shows a hysteresis behavior and is therefore first order by nature, irrespective of the broadness in temperature of the transition.

We note that for films grown on MgO substrates, anti-phase boundaries are inevitable, but perhaps prolonged annealing may help to increase the domain size \cite{Eerenstein03}. For films grown on spinel substrates, one can speculate that perhaps the domain size can also be enlarged by using substrates with smaller lattice mismatch.

\section*{ACKNOWLEDGEMENTS}
We would like to thank T. Mende and C. Becker for their skillful technical assistance. The research of X.H.L was supported by the Max Planck-POSTECH Center for Complex Phase Materials.


\begin{thebibliography}{99}

\bibitem{Groot83} R. A. de Groot, F. M. Mueller, P. G. van Engen, and K. H. J. Buschow, Phys. Rev. Lett. {\bf 50}, 2024 (1983).

\bibitem{Yanase84} A. Yanase, and K. Siratori, J. Phys. Soc. Japan {\bf 53}, 312 (1984).

\bibitem{Groot84} R. A. de Groot, F. M. Mueller, P. G. van Engen, and K. H. J. Buschow, J. Appl. Phys. \textbf{55}, 2151 (1984).

\bibitem{Groot86} R. A. de Groot, and K. H. J. Buschow, J. Magn. Magn. Mater. {\bf 54}, 1377 (1986).

\bibitem{Lind1992} D. M. Lind, S. D. Berry, G. Chern, H. Mathias, and L. R. Testardi, Phys. Rev. B {\bf 45}, 1838 (1992).

\bibitem{Chambers1997} Y. Gao, and S. A. Chambers, J. Crys. Growth {\bf 174}, 446 (1997).

\bibitem{Chambers1997_2} Y. J. Kim, Y. Gao, and S. A. Chambers, Surf. Sci. {\bf 371}, 358 (1997).

\bibitem{Gao97} Y. Gao, Y. J. Kim, S. A. Chambers, and G. Bai, J. Vac. Sci. Technol. A {\bf15}, 332 (1997). 

\bibitem{Chambers1997_4} Y. Gao, Y. J. Kim, and S. A. Chambers, J. Mater. Res. {\bf 13}, 2003 (1998).

\bibitem{Chambers1999} S. A. Chambers, and S. A. Joyce, Surf. Sci. {\bf 420}, 111 (1999).

\bibitem{Chambers2000} S. A. Chambers, S. Thevuthasan, and S. A. Joyce, Surf. Sci. {\bf 450}, L273 (2000).

\bibitem{Chambers2000_2} B. Stanka, W. Hebenstreit, U. Diebold, and S. A. Chambers, Surf. Sci. {\bf 448}, 49 (2000).

\bibitem{ChambersRep2000} S. A. Chambers, Surf. Sci. Rep. {\bf 39}, 105 (2000).

\bibitem{Voogt1997} F. C. Voogt, T. Hibma, P. J. M. Smulders, L. Niesen, and T. Fujii, in Epitaxial Oxide Thin Films III, edited by C-B. Eom, M. E. Hawley, C. M. Foster, J. S. Speck, MRS Symposium Proceedings No. 474 (Materials Research Society, Pittsburgh, 1997), pp. 211-216.

\bibitem{Voogt1998} F. C. Voogt, T. T. M. Palstra, L. Niesen, O. C. Rogojanu, M. A. James, and T. Hibma, Phys. Rev. B \textbf{57}, R8107 (1998).

\bibitem{Voogt1999} F. C. Voogt, T. Fujii, P. J. M. Smulders, L. Niesen, M. A. James, and T. Hibma, Phys. Rev. B \textbf{60}, 11193 (1999).

\bibitem{Bloemen1996} P. J. H. Bloemen, P. A. A. van der Heijden, R. M. Wolf, J. van de Stegge, J. T. Kohlhepp, A. Reinders, R. M. Jungblut, P. J. van der Zaag, and W. J. M. de Jonge, in Epitaxial Oxide Thin Films II, edited by D. K. Fork, T. Shiosaki, J. S. Speck, R. W. Wolf, MRS Symposium Proceedings No. 401 (Materials Research Society, Pittsburgh, 1996), pp. 485-499.

\bibitem{Gaines1997} J. M. Gaines, P. J. H. Bloemen, J. T. Kohlhepp, C. W. T. Bulle-Lieuwma, R. M. Wolf, A. Reinders, R. M. Jungblut, P. A. A. van der Heijden, J. T. W. M. van Eemeren, J. van de Stegge, and W. J. M. de Jonge Surf. Sci. {\bf 373}, 85 (1997).

\bibitem{Gaines1997_2} J. M. Gaines, J. T. Kohlhepp, P. J. H. Bloemen, R. M. Wolf, A. Reinders, and R. M. Jungblut, J. Magn. Magn. Mater. {\bf 165}, 439 (1997).

\bibitem{Anderson_Mg} J. F. Anderson, M. Kuhn, U. Diebold, K. Shaw, P. Stoyanov, and D. Lind Phys. Rev. B \textbf{56}, 9902 (1997).

\bibitem{Heijden95} P. A. A. van der Heijden, J. J. Hammink, P. J. H. Bloemen, R. M. Wolf, M. G. van Opstal, P. J. van der Zaag, and W. J. M. de Jonge, in Magnetic Ultrathin films, Multilayers and Surfaces, edited by A. Fert, H. Fujimori, G. Guntherodt, B. Heinrich, W. F. Egelhoff, Jr., MRS Symposium Proceedings No. 384 (Materials Research Society, Pittsburgh, 1995), pp. 27-32.

\bibitem{Heijden1998} P. A. A. van der Heijden, M. G. van Opstal, C. H. W. Swuste, P. H. J. Bloemen, J. M. Gaines, and W. J. M. de Jonge, J. Magn. Magn. Mater. \textbf{182}, 71 (1998).

\bibitem{Fontijn1997} W. F. J. Fontijn, R. M. Wolf, R. Metselaar, and P. J. van der Zaag, Thin Solid Films \textbf{292}, 270 (1997).

\bibitem{Hibma1999} T. Hibma, F.C. Voogt, L. Niesen, P. A. A. van der Heijden, W. J. M. de Jonge, J. J. T. M. Donkers, and P. J. van der Zaag, J. Appl. Phys. \textbf{85}, 5291 (1999).

\bibitem{Fuji1990} T. Fujii, M. Takano, R. Katano, Y. Bando, and Y. Isozumi, J. Cryst. Growth \textbf{99}, 606 (1990).

\bibitem{Fuji1990_2} T. Fujii, M. Takano, R. Katano, Y. Bando, and Y. Isozumi J. Appl. Phys. \textbf{68}, 1735 (1990).

\bibitem{Fuji1994} T. Fujii, M. Takano, R. Katano, Y. Isozumi, and Y. Bando, J. Magn. Magn. Mater. \textbf{130}, 267 (1994).

\bibitem{Fuji1999} T. Fujii, F. M. F. de Groot, G. A. Sawatzky, F.C. Voogt, T. Hibma, K. Okada, Phys. Rev. B {\bf 59},  3195 (1999).

\bibitem{Mijiritskii2001} A. V. Mijiritskii, and D. O. Boerma, Surf. Sci. {\bf 486}, 73 (2001).

\bibitem{Subagyo2006} A. Subagyo, and K. Sueoka, Jpn. J. Appl. Phys. {\bf 45}, 2255 (2006).

\bibitem{Fonin2005} M. Fonin, R. Pentcheva, Y. S. Dedkov, M. Sperlich, D. V. Vyalikh, M. Scheffler, U. Rudiger, and G. Guntherodt, Phys. Rev. B {\bf 72}, 104436 (2005).

\bibitem{Kiyomura1999} T. Kiyomura, M. Gomi, Y. Maruo and H. Toyoshima, IEEE Trans. Magn. {\bf 35}, 3046 (1999).

\bibitem{Moussy2013} J.-B., Moussy, J. Phys. D: Appl. Phys. {\bf 46}, 143001 (2013).

\bibitem{Verwey39} E. J. W. Verwey, Nature (London) {\bf 144}, 327 (1939).



\bibitem{Eerenstein02} W. Eerenstein, T. T. M. Palstra, T. Hibma, and S. Celotto, Phys. Rev. B {\bf 66}, 201101 (R) (2002).

\bibitem{Kumar06} R. Kumar, M. W. Khan,  J. P. Srivastava, S. K. Arora, R. G. S. Sofin, R. J. Choudhary, and I. V. Shvets, J. Appl. Phys. {\bf 100}, 033703 (2006).

\bibitem{Sena97} S. P. Sena, R. A. Lindley, H. J. Blythe, C. Sauer, M. A. Kafarji, G. A. Gehring, J. Magn. Magn. Mater. {\bf 176}, 111 (1997).

\bibitem{Kale01} S. Kale, S. M. Bhagat, S. E. Lofland, T. Scabarozi, S. B. Ogale, A. Orozco, S. R. Shinde, T. Venkatesan, B. Hannoyer, B. Mercey, and W. Prellier, Phys. Rev. B {\bf 64}, 205413 (2001).

\bibitem{Gong97} G. Q. Gong,  A. Gupta, G. Xiao, W. Qian, and V. P. Dravid, Phys. Rev. B {\bf 56}, 5096 (1997).

\bibitem{Li98} X. W. Li, A. Gupta, G. Xiao, and G. Q. Gong, J. Appl. Phys. {\bf 83}, 7049 (1998).

\bibitem{Arora05} S. K. Arora, R. G. S. Sofin, and I. V. Shvets, Phys. Rev. B {\bf 72}, 134404 (2005).

\bibitem{Arora06} S. K. Arora, R. G. S. Sofin, I. V. Shvets, and M. Luysberg, J. Appl. Phys. {\bf100}, 073908 (2006).

\bibitem{Ramos06} A. V. Ramos, J. B. Moussy, M. J. Guittet, A. M. Bataille, M. G. Soyer, M. Viret, C. Gatel, P. B. Guillemaud, and E. Snoeck, J. Appl. Phys. {\bf 100}, 103902 (2006).

\bibitem{Ziese00} M. Ziese, and H. J. Blythe, J. Phys.: Condens. Matter {\bf 12}, 13 (2000).

\bibitem{Ziese2000} M. Ziese, Phys. Rev. B {\bf 62}, 1044 (2000).

\bibitem{Ziese02} M. Ziese, Rep. Prog. Phys. {\bf 65}, 143 (2002).

\bibitem{Geprags13} S. Geprags, D. Mannix, M. Opel, S. T. B. Goennenwein, and R. Gross, Phys. Rev. B {\bf 88}, 054412 (2013).

\bibitem{Wang13} W. Wang, J. M. Mariot, M. C. Richter, O. Heckmann, W. Ndiaye, P. De Padova, A. Taleb-Ibrahimi, P. Le Fevre, F. Bertran, F. Bondino, E. Magnano, J. Krempasky, P. Blaha, C. Cacho, F. Parmigiani, and K. Hricovini, Phys. Rev. B {\bf 87}, 085118 (2013).

\bibitem{Liu13} M. Liu, J. Hoffman, J. Wang, J. Zhang, B. N. Cheeseman, and A. Bhattacharya, Sci. Rep. 3:1876  DOI: 10.1038/srep01876 (2013); Sci. Rep. 3:3582  DOI: 10.1038/srep03582 (2013).

\bibitem{Naftalis11} N. Naftalis, A. Kaplan, M. Schultz, C. A. F. Vaz, J. A. Moyer, C. H. Ahn, and L. Klein, Phys. Rev. B {\bf 84}, 094441 (2011).

\bibitem{Moussy04} J. B. Moussy, S. Gota, A. Bataille, M. J. Guittet, M. Gautier-Soyer, F. Delille, B. Dieny, F. Ott, T. D. Doan, P. Warin, P. Bayle-Guillemaud, C. Gatel, and E. Snoeck, Phys. Rev. B {\bf 70}, 174448 (2004).

\bibitem{Margulies96} D. T. Margulies, F. T. Parker, F. E. Spada, R. S. Goldman, J. Li, R. Sinclair, and A. E. Berkowitz, Phys. Rev. B {\bf 53}, 9175 (1996).

\bibitem{Orna10} J. Orna, P. A. Algarabel, L. Morellon, J. A. Pardo, J. M. de Teresa, R. Lopez Anton, F. Bartolome, L. M. Garcia, J. Bartolome, J. C. Cezar, and A. Wildes, Phys. Rev. B {\bf 81}, 144420 (2010).

\bibitem{Aragon85} Ricardo Aragon, Douglas J. Buttrey, John P. Shepherd, and Jurgen M. Honig, Phys. Rev. B {\bf 31}, 430 (1985).

\bibitem{Kakol92} Z. Kakol, J. Sabol, J. Stickler, and J. M. Honig, Phys. Rev. B {\bf 46}, 1975 (1992).

\bibitem{Aragon93} R. Aragon, P. M. Gehring, and S. M. Shapiro, Phys. Rev. Lett. {\bf70}, 1635 (1993).

\bibitem{Honig95} J. M. Honig,J. Alloys and Compounds {\bf 229}, 24 (1995).

\bibitem{Brabers98} V. A. M. Brabers, F. Walz, and H. Kronmuller, Phys. Rev. B {\bf 58}, 14163 (1998).

\bibitem{VoogtPhD} F. C. Voogt, Ph.D. thesis, 1998, Materials Science Centre, University of Groningen, the Netherlands.

\bibitem{Huang2013} X. L. Huang, Y. Yang, and J. Ding, Acta Materialia {\bf61}, 548 (2013).

\bibitem{Miles57} P. A. Miles, W. B. Westphal, and A. von Hippel, Rev. Mod. Phys. {\bf29}, 279 (1957). 

\bibitem{Eerensteinprl02} W. Eerenstein, T. T. M. Palstra, S. S. Saxena, and T. Hibma, Phys. Rev. Lett. {\bf88}, 247204 (2002).

\bibitem{Sofin11} R. G. S. Sofin, S. K. Arora, and I. V. Shvets, Phys. Rev. B {\bf83}, 134436 (2011).

\bibitem{Garcia04} J. Garcia, and G. Subias, J. Phys.: Condens. Matter {\bf 16}, R145 (2004).

\bibitem{VanDeVeerdonk} R. J. M. van de Veerdonk, M. A. M. Gijs, P. A. A. van der Heijden, R. M. Wolf, and W. J. M. de Jonge, in  Epitaxial Oxide Thin Films II, edited by D. K. Fork, T. Shiosaki, J. S. Speck, R. W. Wolf, MRS Symposium Proceedings No. 401 (Materials Research Society, Pittsburgh, 1996), pp. 455-460.

\bibitem{Scherrer} DS = 0.9$\lambda$$_{Cu}$/(B$_{struct}$cos$\theta$), where $\lambda$$_{Cu}$ is 1.54056~\textrm{\AA}, B$_{struct}$ describes the structural broadening, which is the difference in integral profile width between a standard (0.006) and the sample to be analyzed, and $\theta$ is the angle of incidence.

\bibitem{Boulch2001} F. Boulch, M. C. Schouler, P. Donnadieu, J. M. Chaix, and E. Djurado, Image Anal Stereol {\bf20}, 157 (2001).

\bibitem{Patterson1939} A. L. Patterson, Phys. Rev. {\bf56}, 978 (1939).

\bibitem{Eerenstein03} W. Eerenstein, T. T. M. Palstra, T. Hibma, and S. Celotto, Phys. Rev. B {\bf68}, 014428 (2003).

\bibitem{Margulies97} D. T. Margulies, F. T. Parker, M. L. Rudee, F. E. Spada, J. N. Chapman, P. R. Aitchison, and A. E. Berkowitz, Phys. Rev. Lett. {\bf79}, 5162 (1997).

\bibitem{Eerensteinjmmm03} W. Eerenstein, L. Kalev, L. Niesen, T. T. M. Palstra, and T. Hibma, J. Magn. Magn. Mater. {\bf258}, 73 (2003).

\bibitem{Hamie12} A. Hamie, Y. Dumont, E. Popova, A. Fouchet, B. W. Fonrose, C. Gatel, E. Chikoidze, J. Scola, B. Berini, and N. Keller, Thin Solid Films {\bf525}, 115 (2012).

\bibitem{Luysberg09} M. Luysberg, R. G. S. Sofin, S. K. Arora, and I. V. Shvets, Phys. Rev. B {\bf80}, 024111 (2009).


\end{thebibliography}
\end{document}